\documentclass[letterpaper]{llncs}

\usepackage{amsmath,times,amssymb,amsfonts,epsfig}
\usepackage[latin1]{inputenc}
\usepackage[english]{babel}
\usepackage{cite}

\def\lengtharrow{50mm}

\def\comp#1{#1^c}

\begin{document} 
 \title{An Asymptotically Optimal RFID Authentication Protocol
Against Relay Attacks}  
\author{Gildas Avoine$^\dagger$ and Aslan Tchamkerten$^\ddagger$} \institute{$^\dagger$Universit\'e
Catholique de Louvain \\
$^\ddagger$TELECOM ParisTech} 
\maketitle \begin{abstract}  
Relay attacks are a major concern for RFID systems: during an authentication
process an adversary transparently relays messages between a verifier and a
remote legitimate prover.

We present an authentication protocol suited for RFID systems. Our solution is
the first that prevents relay attacks without degrading the authentication
security level: it minimizes the probability that the verifier accepts a fake
proof of identity, whether or not a relay attack occurs.

\end{abstract}

{\it{Keywords:}} {authentication protocol, proximity check, relay attack, RFID}
\section{Introduction}

Radio Frequency Identification (RFID) allows to identify objects or
subjects without any physical nor optical contact, using transponders
--- micro-circuits with an antenna --- queried by readers through a
radio frequency channel. This technology is one of the most promising
of this decade and is already widely used in applications such as access cards,
transportation passes, payment cards, and passports. This success is
partly due to the steadily decrease in both size and cost of passive
transponders called \emph{tags}.

The \emph{relay attack}\footnote{Sometimes referred to as {\emph{Mafia fraud}}.} exhibited by Desmedt, Goutier, and
Bengio~\cite{DesmedtGB-1987-crypto} recently became a major issue of concern for RFID
authentication protocols. The adversary pretends to be the legitimate prover by relaying the messages that
are exchanged during the execution of the protocol. This is illustrated through the
following example.

Consider an RFID-based ticket selling machine in a theater. To buy a
ticket, the customer is not required to show his theater pass, an RFID
tag. The customer needs to be close enough to the machine
(verifier) so that the pass (prover) can communicate with it. The pass
can be kept in the customer's pocket during the transaction. Assume
there is a line of customers waiting for a ticket. Bob and Charlie
masterminded the attack. Charlie is in front of the
machine while Bob is far in the queue, close to Alice, the
victim. When the machine initiates the transaction with Charlie's
card, Charlie forwards the received signal to Bob who transmits it to
Alice. The victim's tag automatically answers since a passive RFID tag ---
commonly used for such applications ---
responds without requiring the agreement of its holder. The answer is
then transmitted back from Alice to the machine through Bob and
Charlie who act as relays. The whole communication is transparently
relayed and the attack eventually succeeds: Alice pays Charlie's
ticket.

When it was first introduced in the late eighties, the relay attack
appeared unrealistic. Nowadays, the relay attack is one of the most
effective and feared attacks against RFID systems; it can be easily implemented
since the reader and the tag communicate wirelessly, and it is not
easily detectable by the victim because queried (passive) tags automatically
answer to the requests without agreement of their bearers. Recently,
Halv\'a\v{c} and Rosa~\cite{HlavacR-2007-eprint} noticed that the
standard ISO 14443, related to proximity cards and widely deployed in
biometric passports, can easily be abused by a relay attack due to
the untight timeouts in the communication.

All current authentication protocols that prevent relay attacks perform rather poorly
against an adversary that does not relay messages. They guarantee the same
security level regardless of the adversary's ability to relay messages. 
This may be considered as a weakness, in particular in situations where
relay attacks are hard to perform.

We introduce a new authentication protocol suited for RFID systems with the property that it 
minimizes the false-acceptance probability whether or not a relay attack occurs. In Section~\ref{section:protocol} we present our
protocol. Section~\ref{section:analysis} is devoted to the security analysis. Section~\ref{optimalitty} addresses the optimality of our
solution. In Section ~\ref{section:discussion} we compare our protocol with
related authentication protocols.

\section{Protocol}\label{section:protocol}

\subsection{Protocol requirements and assumptions} \label{requirements}

In the presence of the legitimate prover, the authentication protocol must
guarantee that the verifier always accepts his proof of identity. The protocol
must also prevent an adversary of being falsely identified assuming she can
participate either passively or actively in protocol executions with either or
both the prover and the verifier. This means that the adversary can $1)$
eavesdrop protocol executions between the legitimate prover and the verifier
(passive attack); $2)$ be involved in protocol executions with the verifier and
the legitimate prover separately or simultaneously (active attack). We
assume that neither the prover nor the verifier colludes with the adversary,
i.e., the only information the adversary can obtain is through protocol
executions. Finally, we assume that the legitimate prover and the adversary
never want to get simultaneously authenticated.

Given an integer $N\geq 1$, we consider that the adversary is successful if she is able to impersonate the
legitimate prover within $N$ protocol executions involving either passive
or active attacks. Throughout the paper, $N$ is considered as a fixed constant and, in the RFID
context, may be interpreted as the typical number of authentications the tag can support during its
life.

\subsection{Protocol description}
\label{protocol}
Prior to the protocol execution, the legitimate prover and the verifier agree on a
common secret key $k$ in the form of a binary string of length
\begin{align}
\label{eq:scaling}
\ell_k=2^{n+2}-2
\end{align} for
some integer $n\geq 1$. The protocol
consists of three parts: initialization, authentication, and proximity check. The
initialization and the authentication parts are executed during a ``slow phase'' where no
time measure takes place. The proximity check, instead, involves time measure and is
often referred to as the ``fast phase.''

In addition to $\ell_k$, the protocol involves two positive integers $\ell_a$
and $\ell_b$ whose values will be specified in Section~\ref{section:analysis}.

\subsubsection*{Initialization.}

\begin{figure} \begin{center}
\setlength{\unitlength}{.45bp} \begin{picture}(276,159)
\put(0,20){\includegraphics[scale=.45]{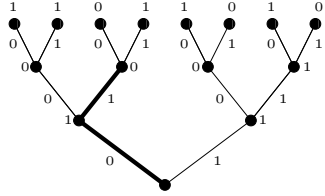}}
\put(130,5){\makebox(0,0){\tiny{}}} \put(85,45){\makebox(0,0){\tiny{$0$}}}
\put(175,45){\makebox(0,0){\tiny{$1$}}} \put(33,97){\makebox(0,0){\tiny{$0$}}}
\put(86,97){\makebox(0,0){\tiny{$1$}}} \put(177,97){\makebox(0,0){\tiny{$0$}}}
\put(230,97){\makebox(0,0){\tiny{$1$}}}

\put(50,79.5){\makebox(0,0){\tiny{$1$}}}
\put(213,79.5){\makebox(0,0){\tiny{$1$}}}

\put(14,124){\makebox(0,0){\tiny{$0$}}}
\put(250,124){\makebox(0,0){\tiny{$1$}}}

\put(105,124){\makebox(0,0){\tiny{$0$}}}
\put(157,124){\makebox(0,0){\tiny{$0$}}}

\put(4,143){\makebox(0,0){\tiny{$0$}}}
\put(41,143){\makebox(0,0){\tiny{$1$}}}

\put(75,143){\makebox(0,0){\tiny{$0$}}}
\put(115,143){\makebox(0,0){\tiny{$1$}}}

\put(147,143){\makebox(0,0){\tiny{$0$}}}
\put(187,143){\makebox(0,0){\tiny{$1$}}}

\put(220,143){\makebox(0,0){\tiny{$0$}}}
\put(260,143){\makebox(0,0){\tiny{$1$}}}

\put(4,174){\makebox(0,0){\tiny{$1$}}}
\put(41,174){\makebox(0,0){\tiny{$1$}}}

\put(75,174){\makebox(0,0){\tiny{$0$}}}
\put(115,174){\makebox(0,0){\tiny{$1$}}}

\put(147,174){\makebox(0,0){\tiny{$1$}}}
\put(187,174){\makebox(0,0){\tiny{$0$}}}

\put(220,174){\makebox(0,0){\tiny{$1$}}}
\put(260,174){\makebox(0,0){\tiny{$0$}}}
\end{picture}
\caption{Decision tree with $n=2$ and $\ell_k=14$. The thick line path in the tree
corresponds to the verifier's challenges $0,1$ and the prover's replies $1,0$.}\label{figure:tree}
\end{center}
\end{figure}

The prover sends a random $\ell_a$-bit string $a$ to the verifier and,
similarly, the prover sends a random $\ell_b$-bit string $b$ to the verifier.
With $a$, $b$, and their common secret key $k$, the verifier and the prover
generate a full binary tree  $\tau(a,b,k)$ of depth $n+1$ as follows (see Fig.~\ref{figure:tree}
for an example). The left and the right edges are labeled $0$ and $1$,
respectively, and each node (except the root) takes the value $0$ or $1$ depending on $a$,
$b$, and $k$. 

The ``tree valued'' function $\tau(a,b,k)$ is a one-to-one function whenever
two of the three variables $a$, $b$, $k$ are kept fixed. (For this to be
possible, $\ell_a$ and $\ell_b$ must be at most equal to $\ell_k$ since the
total number of complete binary trees of depth $n+1$ is equal to
$2^{2^{n+2}-2}=2^{\ell_k}$.) 

\subsubsection*{Authentication.} The prover transmits the $m$ bits
corresponding to the $m$ leftmost leaves, starting from the left. The value of
$m$ will be specified in Section~\ref{section:analysis}. For now, $m$ 
is some value smaller than $2^{n+1}$, the total number of leaves.

\subsubsection*{Proximity check.}
An $n$-round fast bit exchange between the verifier and the prover
proceeds using the tree. The edge and the node values represent the
``verifier's challenges'' and the ``prover's replies,''
respectively. At each step $i\in \{1,2,\ldots,n\}$ the verifier generates a
challenge in the form of a random bit $q_i$ and sends it to the
prover. The prover replies by sending the value of the node in the
tree whose edge path from the root is $q^i=q_1,q_2,\ldots,q_i$. This
reply is denoted by $r_i(q^i)$.

In the example illustrated by Fig.~\ref{figure:tree}, the verifier always
replies $0$ in the second round unless the first and the second challenges are
equal to one in which case the verifier replies $1$, i.e., $r_2(q^2)=0$ for
$q^2\ne 11$  and $r_2(q^2)=1$ for $q^2=11$. 
Finally, for all $i\in \{1,2,\ldots,n\}$, the verifier measures
the time interval between the instant $q_i$ is sent until $r_i(q^i)$ is
received.  

The round-trip time for each challenge-response round
guarantees that the prover is close from the verifier. Hence, a typical
threshold is a value close to $2d/c$ where $d$ denotes the distance from the
verifier to the expected position of the prover and where $c$ denotes the speed
of light.


\begin{figure}[htb]
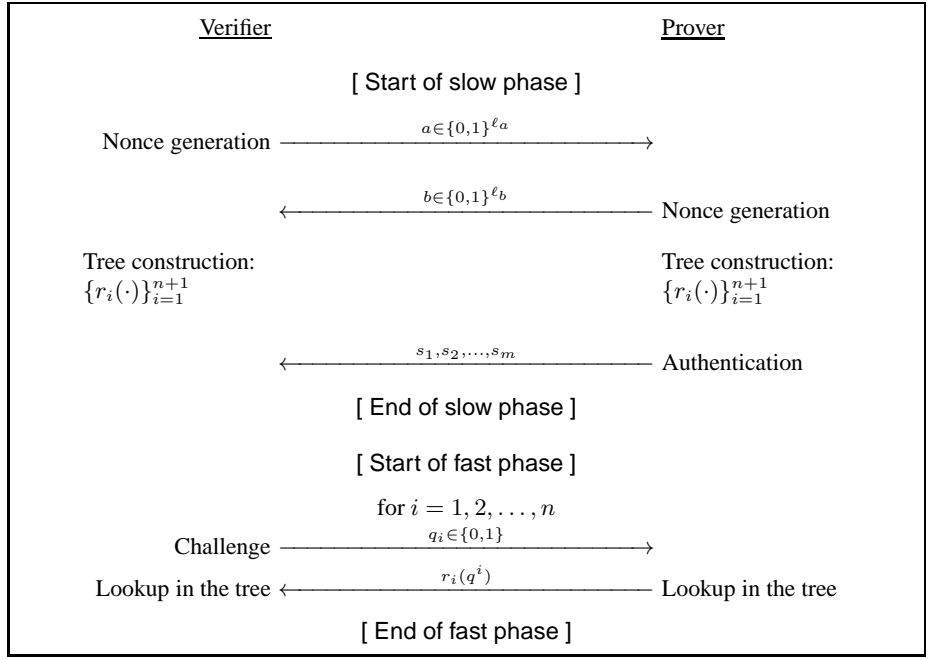

\centering
\fbox{\begin{minipage}{.98\textwidth}
\begin{displaymath}
\begin{array}{rcl}
\text{\underline{Verifier}}& & \text{\underline{Prover}}\\[10pt]
 & \text{\sf [ Start of slow phase ]}&  \\[6pt]
\text{Nonce generation} & \stackrel{a\in\{0,1\}^{\ell_a}}{\hbox to
  \lengtharrow{\rightarrowfill}}    & \\[10pt]
  & \stackrel{b\in\{0,1\}^{\ell_b}}{\hbox to
  \lengtharrow{\leftarrowfill}}    & \text{Nonce generation}\\[10pt]
  \parbox{25mm}{\text{Tree construction:}\\\text{$\{r_i(\cdot)\}_{i=1}^{n+1}$}} & & \parbox{25mm}{\text{Tree construction:}\\\text{$\{r_i(\cdot)\}_{i=1}^{n+1}$}}\\[20pt]
& \stackrel{s_1,s_2,\ldots,s_{{m}}}{\hbox to
  \lengtharrow{\leftarrowfill}}    & \text{Authentication}\\[6pt]
 & \text{\sf [ End of slow phase ]}& \\[10pt]
& \text{\sf [ Start of fast phase ]}& \\[6pt]
& \text{for}\: i=1,2,\ldots,n & \\
\text{Challenge}& \stackrel{q_i\in\{0,1\}}{\hbox to \lengtharrow{\rightarrowfill}} &\\
\text{Lookup in the tree}  & \stackrel{r_i(q^i)}{\hbox to \lengtharrow{\leftarrowfill}}   &  \text{Lookup in the tree}\\[5pt]
 & \text{\sf [ End of fast phase ]}&
  \end{array}
 \end{displaymath}%
\end{minipage}}
\caption{Two-phase distance bounding protocol.}\label{figure:bac}
\end{figure}

\subsubsection*{Final decision.} The verifier accepts the prover's identity
only if the $m$ authentication bits are correct and if the $n$ replies of the
fast phase are correct while meeting the challenge-response time constraint.
The protocol is given in Fig.~\ref{figure:bac}.


\section{Security analysis}
\label{section:analysis}

We are interested in the probability of the event ``over $N$ protocol executions,
the verifier accepts the proof of identity of the attacker at least once.'' To compute
this quantity, we make the following assumption which we discuss below: one protocol execution provides no
information to the attacker about the secret key $k$. As a corollary, the knowledge of
$a$ and $b$ only reveals nothing about the assignment of each node which,
independently, may take the values $0$ or $1$ with probability $1/2$. 

At first, the above assumption may rise some doubts since the $m$
authentication bits and the $n$ bits sent during the fast phase by the prover
depend on the secret key. In practice, however, this assumption may be justified by
arguing that if $m+n$ is much smaller that the size of the key, $\ell_k=2^{n+2}-2$,
one protocol execution reveals almost no information about the secret key.
To be consistent with our assumption, from now on we assume that $m=m(n)=o(\ell_k)$, i.e., that $m$ grows
sub-exponentially with $n$.

To compute the probability of false-authentication, we distinguish two cases depending
on whether during the $N$ protocol executions the adversary acts alone --- i.e.,
without interacting either passively or actively with the legitimate prover ---  or
not. 

\subsection{Attack without involving the legitimate prover}
\label{without}

 We upper and lower bound the probability of
false-acceptance (f-a) as
\begin{align}\label{eqo}
\Pr(\text{f-a}|E)\Pr(E)\leq \Pr(\text{f-a}) \leq 
\Pr(\text{f-a}|E) + \Pr(\comp{E})
\end{align}
where $E$ denotes the event ``over $N$ protocol executions
all trees are different'' and where $E^c$ denotes the complement of $E$.
Conditioned on $E$, the adversary maintains a uniform prior on the secret key $k$
on each protocol execution. Therefore, for each
protocol execution the adversary achieves a probability of success (at best)
equal to $2^{-(m+n)}$, corresponding to random guesses. It follows that
\begin{align}\label{nor1}
\Pr(\text{f-a}|E)=N\cdot 2^{-(m+n)}+o(2^{-(m+n)})\quad \quad
(n\rightarrow \infty)\;.
\end{align}
The computation of $\Pr(E^c)$ refers to the birthday paradox. By letting
$\ell_a=m+n$, a standard
calculation reveals that\footnote{The bound
\eqref{bpar} is achieved if $\ell_b$ is kept fix during the $N$ protocol
executions.}
\begin{align}\label{bpar}
\Pr(E^c)\leq\frac{N(N-1)}{2^{m+n+1}}+O(2^{-2(m+n)})\quad \quad(n\rightarrow
\infty)\;.
\end{align}
From \eqref{eqo},\eqref{nor1}, and \eqref{bpar}
we get
\begin{align}\label{saispas}
\Pr(\text{f-a})=\Theta(2^{-(m+n)})\quad \quad (n\rightarrow \infty)\;.
\end{align}
\subsection{Attack involving the legitimate prover}
We distinguish two sub-cases, depending on whether the adversary can or cannot relay messages.
\subsubsection{With relay.} In this case, the adversary can execute man-in-the-middle attacks to pass the
authentication step for each of the $N$ protocol executions; the adversary initiates the protocol with the
verifier and relays the nonces $a$, $b$, and the authentication
string $s_1,s_2,\ldots,s_{m}$. However, to succeed the adversary must pass the
proximity check. We compute the probability of false-acceptance
(f-a) assuming the adversary passed the authentication step. Similarly as in
\eqref{eqo}, we upper and lower
bound the probability of false-acceptance as
\begin{align}
\Pr(\text{f-a}|E_b)\Pr(E_b)\leq \Pr(\text{f-a}) \leq 
\Pr(\text{f-a}|E_b) + \Pr(\comp{E_b})\label{eq1}
\end{align}
where $E_b$ denotes the event ``over $N$ protocol executions
all $b$ nonces are different.''

We first compute $\Pr(\text{f-a}|E_b)$. Because of the time constraint, the
adversary cannot relay information between the verifier and the prover during
the fast phase. This means that the adversary's reply at time $i$ must be
independent of the verifier's challenge at time $i$, for any $i\in
\{1,2,\ldots,n\}$. However, because there is no time measure before the fast
phase, the adversary can query the legitimate prover with a sequence of
challenges  $\tilde{q}^n$, hoping these will
correspond to the challenges ${q}^n$ provided by the verifier during the fast
phase. Because $q^n$ and $\tilde{q}^n$ are independently chosen, the
probability of passing the proximity check is the same for any $\tilde{q}^n$.
Hence, without loss of generality, we assume that the adversary has access to
the $r_i(\tilde{q}^i)$'s for $\tilde{q}^n=(0,0,\ldots,0)\triangleq 0^n$. The
adversary is then  successful only if $r_i(0^i)=r_i(q^i)$ for all $i\in
\{1,2,\ldots,n\}$. For conciseness, from now on we write $r_i$ for $r_i({q}^i)$
and $\tilde{r}_i$ for $r_i(\tilde{q}^i)$.

Letting $t$ be the first time $i\geq 1$ when $q_i=1$, we have that $\tilde{r}_i=r_i$ for $i\in
\{1,2,\ldots,t-1\}$, and $\tilde{r}_i=r_i$ with probability $1/2$ for $i\in
\{t,t+1,\ldots, n\}$. Therefore, letting $r^n\triangleq
r_1,r_2,\ldots,r_n$, the probability of a successful attack over one
particular protocol execution can be computed as
\begin{align*}
 \Pr(\tilde{r}^n={r}^n)&=\sum_{i=1}^{n}
\Pr(\tilde{r}^n={r}^n|t=i)\Pr(t=i)\nonumber\\
&+\Pr(\tilde{r}^n={r}^n|q^n=0^n)\Pr(q^n=0^n) \nonumber\\
&=\sum_{i=1}^{n}2^{-(n-i+1)}2^{-i}+2^{-n}\nonumber \\ &=2^{-n}(n/2+1)
\end{align*}
and we get
\begin{align}
 \label{gildas}
 \Pr(\text{f-a}|E_b)= 2^{-n+o(1)}\quad \quad (n\rightarrow \infty) \;.
\end{align}
Similarly as in \eqref{bpar} we have
\begin{align}
\Pr(\comp{E_b}) \leq \frac{N(N-1)}{2^{\ell_b+1}}+O(2^{-2\cdot\ell_b})\label{eq2}\quad
(\ell_b\rightarrow \infty)\;.
\end{align}
By taking $\ell_b\geq n$, from \eqref{eq1}, \eqref{gildas}, and \eqref{eq2} the highest probability of false-acceptance that can be attained by an
adversary who can relay messages satisfies
\begin{align}\label{der}
\Pr(\text{f-a}) &  =2^{-n(1+o(1))}\quad \quad  (n\rightarrow \infty)\;.
\end{align}
\subsubsection{Without relay}
As one might observe, the security analysis in the above case ``with relay'' never uses the nonce
$a$. 
Suppose the adversary cannot relay signals. Without the nonce $a$,  the
adversary can easily pass the authentication step by first obtaining the nonce
$b$ and the corresponding authentication string from the legitimate prover, then by presenting those to the
verifier. The security is then based only on the proximity check. Instead, with
a nonce $a$, this attack is less likely to succeed. Indeed, one can readily
see that with $\ell_a=m+n$ as in Section \ref{without}, the probability of false-acceptance is as small as in the case
of attacks without legitimate prover and is
given by \eqref{saispas}.
\section{Optimality of the proposed protocol} \label{optimalitty}
We discuss the optimality of the proposed protocol by restricting our attention
to bit exchange protocols that satisfy the following general properties:
\begin{itemize}
\item[$\bullet$]
The verifier and the legitimate prover share a common secret in the form of a
bit string of length $\ell_k$.
\item[$\bullet$]
The verifier always accepts the proof of identity of a legitimate prover.
\item[$\bullet$]
Neither the verifier nor the legitimate prover collude with the adversary. 
\end{itemize}

Consider an authentication protocol that satisfies the above conditions. Among
the bits sent by the prover during the execution of the protocol, some depend
on the common secret, and some do not. If $m+n$ denotes the
number of secret dependent bits, the false-acceptance probability (per adversary
trial) of the protocol is at best $$2^{-(m+n)}$$ regardless of the type of
attack. 

To overcome relay attacks, it is necessary that the verifier has a means to
determine whether the prover is close to him --- in our case the
time measure. If $n$ denotes the number of key dependent bits sent by the prover upon which
the verifier evaluates his proximity,  the probability of false-acceptance (per
adversary trial) in the presence of relay attacks is at best $$2^{-n}\;.$$

In light of \eqref{saispas} and \eqref{der}, our protocol is asymptotically
optimal  in the sense that the exponential rate at which  the false-acceptance
probability goes to zero as $m$ and $n$ tend to infinity is the best one can achieve
among all protocols with the same parameters.

\section{Discussion}\label{section:discussion}

Brands and Chaum~\cite{BrandsC-1993-eurocrypt} were the first to propose an
authentication protocol using the idea of a proximity check (or distance
bounding) between the prover and the verifier.\footnote{This idea was
originally developed in an earlier work from Beth and
Desmedt~\cite{BethD-1990-crypto}.} This protocol, similarly to ours, uses a
proximity check in the form of rapid exchanges of challenges and responses
between the verifier and the prover. After this phase, the prover authenticates
himself by sending an $m$ bit signature of all sent and received bits --- the
value of $m$ is not specified. 

There are two possible attacks. The adversary can first query the legitimate
prover with a particular sequence of challenges. Whenever the verifier picks
the same sequence of challenges, the adversary succeeds. The other attack
consists in guessing the final signature. The probability of
false-acceptance over $N$ protocol executions is thus approximatively $N\cdot
2^{-\min\{m,n\}}$, $n$ being the number of responses provided during the fast
phase.  Since the only key dependent bits are the $m$
ones of the signature, this protocol is optimal if $m\leq n$ and suboptimal
otherwise. 

Note that, although Brands and Chaum's protocol may be optimal, depending on the choice of
the parameters $m$ and $n$, once the number of fast
phase rounds is fixed, our protocol achieves a much lower probability of
false-authentication in the non-relay case --- and the same in the relay case.


All the subsequently published protocols
\cite{HanckeK-2005-securecomm,MunillaOP-2006-rfidsec,ReidGTS-2006-eprint,MeadowsPPCS-2007-book,CapkunBH-2003-sasn,TuP-2007-rfidtechnology,CapkunH-2006-ieee,SingeleeP-2007-esas}
that prevent relay attacks, while having other features in terms of their
complexity (computations, memory, amount of information exchanged) and their
functionalities (mutual authentication, resistance to noise, resistance to
colluding attacks), attain a probability of false-acceptance at best equal to
the one of Brands and Chaum in both the cases with and without relay. Part of
the reason is because authentication and proximity check are performed
on the basis of the same bits. In our case instead, the bits
sent during the authentication and during the proximity check differ. This main
feature allows us to dramatically reduce the probability of false-acceptance in
situations where relays are not implementable, yet active attacks are
possible. 


We now compare our protocol with Hancke and Kuhn's
\cite{HanckeK-2005-securecomm} since the structures of the fast phases are
related. In Hancke and Kuhn's protocol, two registers $x_1,x_2,\ldots,x_n$ and
$y_1,y_2,\ldots,y_n$ are generated according to the secret key $k$ and the
random nonces $a$ and $b$. For each round $i$ of the fast phase, the legitimate
prover replies $x_i$ or $y_i$ depending on whether the verifier's challenge
$q_i$ is equal to zero or one.  The difference with our protocol is that the
response at time $i$ depends only on the \emph{current} challenge $q_i$ and not
on the \emph{past} challenges $(q_1,q_2,\ldots,q_{i-1})$, i.e.,
$r_i(q^i)=r_i(q_i)$.\footnote{The two registers can be seen as forming a
decision tree where, at any level, each node value depends only on whether it
is issued by a left or a right branch.} Because the adversary can query the
prover during the slow phase, she can obtain the equivalent of an entire
register. As a consequence, the probability of false-acceptance over $N$
protocol executions is approximatively $N\cdot (\frac{3}{4})^n$, which is
significantly higher than for our protocol --- both with and without relay.

We end this section with a practical consideration on our protocol.
Interestingly perhaps, even if it gets interrupted during the fast
phase, the verifier may still provide some reliable decision on whether to
accept or to reject the prover's identity. (Of course, the probability of
false-acceptance will depend on how many replies the verifier obtained.) This
may be useful in situations where fast authentications are required --- e.g.,
for toll gates on highways ---  since it allows the verifier to take a decision
even if the protocol did not end properly.   

\section{Concluding remarks}


The main contribution of this paper consists in a an authentication protocol that is asymptotically
optimal in terms of probability of false-acceptance both in the relay and non-relay cases, in
contrast with previous protocols.

The performance of the protocol, however, comes at the expense of additional storage capabilities in
order to compute the entire decision tree before executing the fast phase. This makes the protocol
mostly suitable in applications where the number of fast phase rounds can be made small --- for
instance, in situations where relay attacks are expected to occur rarely. Numerically, taking $n=11$
for instance, requires a $1$KByte memory. Most RFID tags devoted to secure applications offer this
value --- the common NXP Mifare Classic Standard tag provides a $1$KByte memory and
ICAO-compliant electronic passports embed an at least 30KByte memory tag.

Finally, we note that several other optimality criteria may be considered in addition to the
one proposed in Section \ref{optimalitty}. An interesting direction to pursue might be, given the size of
the secret key, to seek the tradeoff between the probabilities of
false-acceptance with and without relay.

\bibliographystyle{plain}
\bibliography{bibliography}

\end{document}